\documentclass{ringb99}
\usepackage{graphics}
\def\etal{{\it et al.\/}}
\def\eg{{\it e.g.}}

\def\cf{{\it cf.}}
\def\ltw{\>\hbox{\lower.25em\hbox{$\buildrel <\over\sim$}}\>}
\def\gtw{\>\hbox{\lower.25em\hbox{$\buildrel >\over\sim$}}\>}
\begin{document}

\title{Magnetic Fields in Clusters:  Theory {\it vs.} Observations}

\author{Jean Eilek}
\institute{Physics Department, New Mexico Tech, Socorro NM 87801, USA} 

\date{draft:  \today}

\maketitle

\begin{abstract}

It is now well established that the plasma in galaxy clusters is
magnetized.  In some cases, at least, the field is strong enough to be
dynamically important.  Perhaps this is the time to move past simple
detection experiments, and to work towards a general understanding of
the strength, structure and maintenance of the cluster field. 

\end{abstract}

\section{Introduction}

I am not aware of any astrophysical plasma in which the magnetic field
is clearly unimportant.  Space plasmas, which we can observe {\it in
situ}, are commonly magnetized; in some cases the magnetic field
and associated currents dominate the plasma dynamics.  The
interstellar medium (ISM) in our galaxy is magnetized, at
approximately an equipartion level.  The diffuse plasma in radio
galaxies is obviously magnetized, probably at dynamically important
levels.   

Until recently, magnetic fields have generally been ignored in
dynamical models of the intracluster plasma medium (ICM).  However,
this is changing as data accumulates -- and the discussion at this
conference is a case in point.  It seems very likely that the ICM is
magnetized in every cluster -- just as we believe the interstellar
medium (ISM) is
magnetized in every galaxy.  I suspect the dynamics of the magnetized
ICM are in many ways similar to the dynamics of the magnetized ISM.
The magnetized plasmas in both systems are probably subject to strong
effects of  cluster, or galactic, weather ({\it e.g.}, Binney 1998,
Zweibel \& Heiles 1997).  That is, seed fields are probably provided
to the ISM by stellar ejecta; and to the ICM by ejecta from both
active and ``normal'' galaxies (Kronberg {\it et al} 1999, V\"olk \&
Aronyan 1999).  The seed fields are enhanced and shaped by turbulence,
flows and dissipation, a complex process which results in the fields
we see at present.

Up to now observational work on cluster fields has mainly been aimed
at detection:  can we find  any evidence of magnetic fields in any
clusters?  The answer to that seems clearly to be ``yes'', as this
meeting has demonstrated.  It may be time to change our focus.  We
know that the ICM is magnetized;  we should concentrate now
exploring the cluster field in more detail, both observationally and
theoretically.  In this paper I will try to convey a rather personal
impression of what we know about magnetized plasmas from other sources
(laboratory and space data, simulations, theory), and to consider how
this information can be used to understand magnetic fields in
clusters. 

\section{What Does the Data Tell Us?}

We have two ways to detect magnetic fields in the cluster plasma. 
We can detect diffuse synchrotron emission (from radio haloes and
relics), or we can detect Faraday rotation from individual radio
galaxies, either behind the cluster or embedded in it.

\subsection{Overview of the Observations}

We have three possible observational approaches to measuring the
magnetic field in a cluster of galaxies. 

{\bf Radio Haloes and Relics}.  Some clusters show diffuse synchrotron
emission which is associated with the ICM rather than with discrete
radio galaxies within the  clusters ({\it e.g.} Feretti \& Giovannini
1996, Feretti 1999).   This emission can exist
throughout the cluster (haloes), or be peripheral (relics).  In either
case, this is direct evidence of  intracluster B fields.  
Minimum-energy or minimum-pressure arguments, sometimes combined with
X-ray inverse Compton detections or limits, suggest $B \ltw 1 \mu$G is
typical on large scales (comparable to the size of the cluster core,
or the cluster itself). 

{\bf Faraday Rotation of Background Sources}.  Comparison of Faraday
signals from cluster and non-cluster sources allows a statistical test
of cluster magnetic fields.  Current studies do show an excess rotation
measure from sources in or behind rich clusters (Kim {\it et al} 1991,
Clarke 1999).  Modelling this data in terms of a ``typical'' ICM
density and scale, and a cell size $\sim 10$ kpc, again suggests a
microgauss-level field is typical of the cluster on moderate to large
scales.

{\bf Faraday Rotation of Embedded Sources}.  Another approach 
uses high-resolution images of radio sources within the cluster.  As
this is the closest to my own research, I will say a bit more about
it. Radio imaging gives us our only  evidence on the structure of the
magnetic field.  Most studies have been of cluster-center sources, but a 
few studies of non-central sources now exist ({\it e.g.} Feretti {\it
et al} 1999).  In Eilek \& Owen (1999), we present new Faraday data on
two large, tailed radio sources (3C75, 3C465) in the centers of two
clusters (A400, A2634 respectively) which have only 
weak cooling cores.  We also collect published results of other imaged
Faraday data on cluster-center sources (which have studied smaller
radio sources in strong cooling cores). 
These results are summarized in Table 1, which lists
the radio source, its radial extent, the
characteristic scale of its measured rotation measure, the mass
accretion rate inferred from simple cooling-flow analysis, the mean
magnetic field along the line of sight inferred from Faraday data, and
the ratio of magnetic pressure to gas pressure.  This data allows us
to make a general statement:  the magnetic field is dynamically
significant in the central regions of these clusters.  The typical central 
field strength $\sim 10-30 \mu$G.  Comparison to pressures derived
from X-ray data shows that $B^2 / 8 \pi \sim n k_B T$ is typical for
the cluster-center sample.  We can also use the images to estimate
coherence lengths for the rotation measure.  The general trend we see,
is that cthe luster-core rotation measure has a characteristic length
scale $\sim$ 10 kpc.  This is our only direct information so far on
the coherence length of the cluster field.

\begin{table}
\begin{tabular}{lccccc}
\multicolumn{6}{c}{{\bf Table 1}. Cluster-Center Magnetic 
Fields$^{\rm  a}$}\\
\hline\hline
Source &$ L_{RS}$ & $L_{RM}$ & $\dot M$ & $\langle B_{\parallel} \rangle$
     & $ p_B / p_g$ \\
  & kpc  & kpc & $M_{\sun}$/yr & $\mu$G 
\\
\hline
M87              &  3  & 1  & 10  & 50 & 0.7 
\\ 
A2199            & 4  & 3 & 150 & 60 & 5.0 
\\
A1795            & 7 & 3  & 400 & 10 & .09 
\\
A2052            & 8 & 2  & 90  & 40 & 4.5 
\\
A2029            &  10 & 3 & 370 & 75 & 13
\\
A4059            &  10 & 3 &  120 & 25 & 1.8 
\\
Hydra A          &  50 & 5 & 600 & 60 & 13 
\\
Cyg A            & 70 & 10  & 200 & 40 & 4.5 
\\
A 400            &  100 & 10 & $< 1$ & 7 & 0.7 
\\
A 2634           &   140& 20  & $< 1$ & 5 & 0.4 
\\
 \hline\hline
\end{tabular}
\begin{list}{}{} 
\item[$^{\rm a}$]From Eilek \& Owen 1999, and references therein.
\end{list}
 \end{table}

\subsection{What is the Nature of the Cluster Field?}

Collecting all the evidence, I think there is a strong case for
magnetic fields being common in all clusters at (at least) modest
levels.  Both the radio haloes and the Faraday rotation of background
sources are more or less consistent with a microgauss level field
existing throughout most clusters.\footnote{Why don't we detect radio
haloes from all clusters? I suspect it is a question of sensitivity.
The ICM is very likely magnetized in nearly all 
clusters. The galaxy ejecta that carry seed fields should also carry
``seed'' cosmic rays. Could it be the case, then, that we are seeing
only the brightest few haloes with our present observations?}  A
field at this level is not dynamically important at typical ICM densities
and temperatures.  We also know, however, that stronger fields (tens
of $\mu$G) exist in cluster centers, where we detect them in front of
central radio galaxies.  These fields are dynamically important.
There are also hints of stronger fields associated with non-central
radio galaxies.  We do not yet know whether these stronger fields are
typical of the diffuse ICM in all clusters, or whether they are
amplified by energy input from the radio galaxies. 

What is the next step?	
We would like to use these observations to answer fundamental
questions.  What is the strength and structure of the field?  What is
its role in, and relation to, the dynamics of the cluster gas?  The
observations we can make are both tantalizing and frustrating.  They
cannot answer these qustions directly, but can provide important
clues.  We should look to other information that we have on magnetized plasmas.

\section{There Are No Uniform Magnetic Fields}

One is often tempted to interpret an observation in terms of the
simplest possible model.  For the case of magnetic fields, the
very simplest model is a uniform field throught the cluster; the next
simplest is  the old idea of random, Gaussian ``cells'' of field.
We should be very cautious here.  These seems to be no plasma
(whether astrophysical or simulated) in which the field is truly
uniform, nor in which the field structure can be described by simple,
space-filling cells.  The magnetic field in a magnetized plasma is
more likely to be bunched into elongated, high-field regions:  we can
call  these ``filaments'' or ``flux ropes''. 

\subsection{Evidence for Filamented Fields}

The evidence comes from several areas. 

{\bf Simulations of MHD Turbulence} find that intermittency is common:
the magnetic field is organized into localized, often elongated,
high-field regions {\it e.g.}, Menguzzi {\it et al} 1981, Nordlund
{\it et al} 1992, Miller {\it etal} 1996, Kinney {\it et al} 1995). 
This appears similar to the small-scale vortex intermittency seen in
hydrodynamical turbulence;  vorticity in fluid flows sometimes behaves
similarly  to magnetic fields in MHD flows.   The characteristic
intermittency scale is  thought to be similar to the turbulent
dissipation scale (but this conculsion may still be sensitive to
numerical resolution).  

{\bf Space Plasma Data} is measured {\it in situ}.  These measurements
again find direct evidence that the field is often bunched into flux
ropes. These are apparent in the flux transfer events that are part of
patchy reconnection at planetary magnetopauses, and they have been
detected in interplanetary space, associated with coronal mass
ejections. (The compilation edited by Russell \etal\ 1990 is a good
source of references to this area.)  In addition, optical and X-ray
images of the sun make it clear that solar plasma is filamented;  this
is almost certainly due to magnetic fields.\footnote{In many ways the
sun is a poor analogy to cluster magnetic fields, due to the very
heavy ball of gas in which the field lines are ``footed''.  However I
suspect that the field dynamics on small scales are probably similar,
although one must take into consideration the more ordered turbulent,
convective flow that exists on the sun.}

{\bf The Magnetized ISM} displays rich substructure and
filamentation. Dramatic examples include the filamentation apparent in
HI images (Hartmann \& Burton 1997), or recent radio images of the
galactic center (Lang {\it et al} 1999).  Once again, we do not have
direct evidence that the filaments are magnetically created, but such
seems very likely to be the case. 

{\bf Extragalactic Radio Sources} commonly show filamentation in their
synchrotron emission.  A particularly beautiful example is the new
image of the M87 halo (Owen, Eilek \& Kassim 1999, also Owen 1999).
These are almost certainly high-field regions; polarization data when
available shows that such filaments are aligned with the magnetic
field.

\subsection{Consequences of Filamented Fields}

Inhomogeneous fields affect ``simple'' interpretation of observations
in several ways; we must be cautions. 

 {\bf Minimum pressure or equipartion} estimates are well known
to be sensitive to the inhomogeneity of the source.  In following the
usual applications (for instance, that of Pacholczyk 1970), we must
remember that two filling factors are needed.  One is the fraction of
the volume filled by relativistic particles, which may of course be
less than unity.  The second is the fraction of the volume filled by
relativistic particles that is {\it also} filled with magnetic field
(denoted by $\phi$ in the standard notation).  Incomplete volume
filling leads to a {\it higher} magnetic field value, derived from
these standard arguments, than would be the case for a homogeneous
source. 

 Interpretation of {\bf inverse Compton} (IC) X-ray detections or
limits is also sensitive to the degree of mixing of the relativistic
electrons and the magnetic field.  Incomplete mixing (such as magnetic
filamentation) will lead to a lower ratio of synchrotron to IC
emission than would apply to a homogeneous source. Conversely, an
observation of a fixed IC/synchrotron ratio requires a higher magnetic
field value if the source is incompletely filled.  

{\bf Particle aging} due to synchrotron losses is also sensitive
to magnetic field inhomogeneity.  Particles lose energy more rapidly,
and radiate at higher frequencies, in a strong field.  One consequence
of this is that the high-frequency decay of the synchrotron spectrum
can be dramatically slowed down if the particles spend much of the
time in weak-field regions (Eilek, Melrose \& Walker 1997).  Applied
to interpretation of data, one must be aware that radio sources can be
much older than they look.

Finally, interpretation of {\bf synchrotron spectra} must be
done with care if the magnetic field is inhomogeneous.  The usual
analysis assumes a uniform field, and interprets the photon spectrum
as a direct measure of the particle energy distribution.  In reality,
however, the photon spectrum is a convolution of the field and
particle distributions.  One example is that a power-law distribution
of magnetic field strength, combined with a peaked particle
energy distribution, can produce a power-law photon spectrum
(Eilek \& Arendt 1996).  It is quite likely that observed spectra are
telling us as much about the field distribution as about the
particles.

\subsection{Nature of the Filaments}

What can we say about the small-scale structure of these magnetic
filaments?  I think our best evidence comes from space plasmas, where
{\it in situ} field measurements are possible.  The filaments tend to
be structured as a {\it flux rope}.  That is, a twisted (helical)
magnetic field structure, in which the field lies along the axis in
the center of the tube, and becomes increasingly helical going away
from the axis.   Such flux ropes can be confined by external pressure,
or can be partly or fully self-confining, depending on the exact
structure of the field.  The latter is particularly nice, as it allows
strong fields to be self-organized into filamentary structures without
the need for external pressure confinement.

What is the origin of these structures?  Fluid flow offers analogs.
Consider is a tornado:  a strong vortex line which
forms due to the combination of shear and convergeance in the local
atmosphere. As vorticity and magnetic field are often useful analogs of
each other,  one would expect magnetic flux ropes to arise from
similarly favorable flow fields.  (One MHD example might be the solar
convective 
zone, where converging flows accumulate the magnetic field into
strong, localized, low-density filaments.   Add a bit of 
circulation and 
one can imagine a twisted flux rope arising.) One can also look to
space plasma analogs.  Flux rope formation associated with flux
transfer events is thought to be due to plasma instabilities,
particularly reconnection, again in the presence of some twist.
Tearing instabilities lead to magnetic island 
formation in two-dimensional problems (such as the neutral sheet in a
reconnection situation), and thus connect to helical flux ropes in three
dimensions.  

In other domains, MHD simulations  find that passive magnetic fields
(weak compared to flow energy density) in sheared flows collect into
filaments (such as the radio galaxy simulations of  Clarke 
1996).   Laboratory evidence shows that strong fields (compared to
plasma energy densities) can organize themselves into force-free
structures ({\it e.g.} Taylor 1986); in cylindrical geometry these have
the general flux-rope structure described above.  Thus, it seems that
we can generally think of magnetic fields as filamented, probably on a
ranges of scales, whether the field is weak or strong.

\section{Large Scale Structure of the Magnetic Field}

The existence of small-scale filaments does not preclude an organized
large-scale structure.  An obvious example is the sun:  the surface
field is strongly inhomogeneous and filamented, and yet shows a simple
dipolar structure when measured on larger scales.  What can we say
about the large-scale structure of the cluster field?

\subsection{Observational Evidence}

We have little direct information as yet.  We know that the magnitude
of the rotation measure seems to drop going away from cluster center;
this is apparent in the statistical data of Clarke (1999), and also in
the two large-scale sources measured by Eilek \& Owen (1999).  We also
know that large-scale radio haloes have a finite size, which is no
larger than the cluster size. 

Both of these observations suggest that the field strength decreases
with distance from the cluster center (except possibly clusters with
in offset radio relics, which may have an enhanced magnetic field near
their periphery).  However, both observations must be interpreted with
care.  The observed decay of rotation measure must be due in part to
the decrease of gas density, as well as to the field structure.
Similarly, the structure of a radio halo is determined by the
relativistic particle density as well as the field strength. 
I am not aware of any work at this point which can separate out the
effects of the field and the particles in these observations. 

\subsection{Theoretical Arguments}

We can make one general statement:  cluster fields must involve a
dynamo.  For our purposes, I define a dynamo (interpreted
loosely) as anything that plasma flows can do to a magnetic field.
The most common application involves the inductive enhancement of a
weak seed field to interesting levels.  
(This comes from the $\bf{v} \times \bf{B}$ term in the induction
equation, given for instance in Garasi \& Eilek 1999)  As a general rule,  
flows can amplify a field up to a state of dynamic balance, $B^2 \ltw
4 \pi \rho v^2$.  However, the 
rate of growth and the final field level depend on details of the flow
field, particularly turbulent and dissipative effects. In addition,
some situations (such as flux ropes in the sun) result in the field on
small scales coming into balance with the static pressure:  $B^2
/ 8 \pi \sim n k T$.   This results in a higher field that that
produced by the usual dynamic balance estimate, if the flows
controlling the dynamo are subsonic. Figure 1  illustrates some of the
possible branch points in turbulent and/or advective dynamos.  

\begin{figure}
 \resizebox{\hsize}{!}{\includegraphics{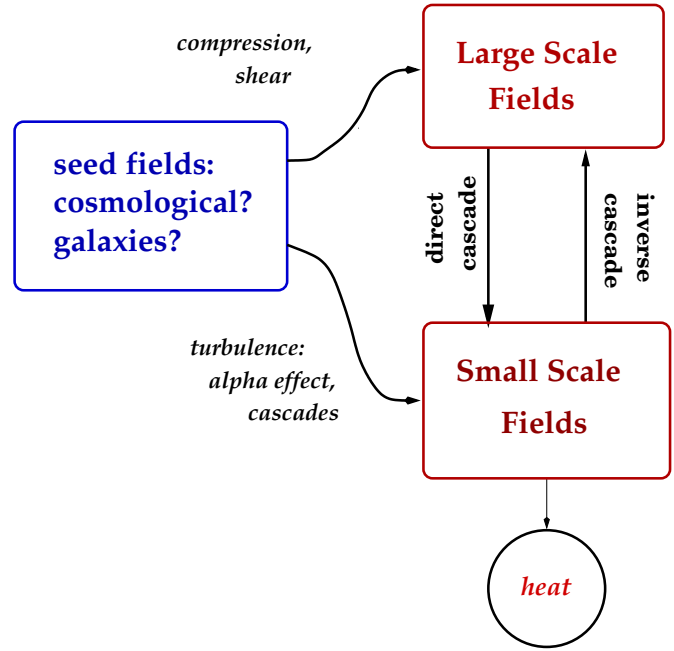}}
\caption[]{Schematic flow chart for a cluster dynamo.  The key point
is that plasma flows can support both large-scale and small-scale
magnetic fields;  the nature of the resultant field depends, of
course, on the detailed properties of the flow.  Beck \etal\ (1996)
has a similar figure and a detailed discussion of galactic dynamos.}
\end{figure}

What can dynamo theory tell us about the cluster field?  We can, in
principle, inquire about the strength and structure of the field. 

\subsubsection{Field Strength}

Several authors have suggested that turbulent dynamos maintain a low
level magnetic field throughout the cluster ({\it e.g.} Ruzmaikin {\it
et al} 1989, Goldman \& Rephaeli 1991, De Young 1992). These models
generally address only the local field strength (from the energy
balance relation given above), and assume the field is disordered with
small-scale tangling.  The field strength depends directly on the
turbulent velocity assumed:  $v_t = 400$km/s gives $B = 6 \mu$G if the
plasma density is $10^{-3}$cm$^{-3}$. 

An important related question is what drives the turbulence that these
models require to maintain the field.  Earlier work argued that galaxy
driving in a relaxed cluster could maintain interesting levels of
turbulence, and magnetic field, throughout 
the cluster. However, more recent work disagrees.  There now seems to
be a general agreement that galaxy driving cannot maintain the strong
turbulence, (probably transonic;  {\it e.g.} Baum 1992, or Keel {\it
et al} 1996) and strong magnetic fields, seen in cluster cores.  It
may not even be able to maintain a microgauss level field.  I have done
calculations for turbulent driving in a rich, Coma-type cluster, and
find turbulent velocities of a few tens of km/s, which maintain fields
at only a fraction of a microgauss.  

This difficulty leaves one with two options.  One can argue that the
stronger fields discussed at this meeting are atypical, perhaps
maintained only locally by energy input from an embedded radio
galaxy.  Alternatively, ongoing evolution of the cluster gravitational
potential (otherwise known as cluster mergers) can provide a stronger
energy input, in principle maintaining stronger turbulent and field
levels.  Norman \& Bryan (1998) indeed found turbulent velocities
between 1/4 and 1/2 of the virial velocity were maintained for
significant times in one set of merger simulations.

\subsubsection{Field Structure}

Less work has been done on the spatial structure of the magnetic
field.  

Several authors have noted that the compression associated with
convergent cooling flows will amplify an initial seed field towards
the cluster center.  For a field which remains purely turbulent and
simply advected, one would expect $B \propto \rho^{3/2}$ (this would
be the case if fluid turbulence or reconnection maintained a locally
turbulent field during the inflow).  An alternative picture is that of
Soker \& Sarazin (1990), who proposed that the inflow stretches an
initally disordered field radially.  They find the field varies as $B
\propto 1 / r^2$ in this case.  Such arguments also assume a weak,
passive field; when the field becomes strong it will affect the flow
and break these scaling laws.  Numerical simulations (such as Dolag,
Bartlemann \& Lesch 1999, Garasi \& Eilek 1999, Roettinger 1999) will
be required to determine the final answer here.   

What field structure td simple turbulent dynamo models predict?
One expects that the turbulently maintained field will be strong where
the turbulence in strong.  Locally strong turbulence might be the
reason why some haloes, or relics, do not share the symmetry of the
cluster in which they sit.  I have not seen this addressed in
published work; again, ongoing numerical simulations have the
potential to address this.       

The behavior of turbulent dynamos has the potential to be much richer
than these simple arguments.  In particular, laboratory plasmas show a
capacity for {\it self-organization}.  If the system has the proper
initial asymmetry (helicity), the magnetic field can undergo a
 ``relaxation'' process to a final state which displays more order on
large scales than the initial state had (Taylor, 1986).  I have
considered one possible final state for clusters, namely, nested
toroidal flux 
tubes.  This solution predicts that the magnetic field increases
outward from a central minimum, and then decays as $B \propto 1/r$.
It is not yet clear whether these formal solutions (derived from
laboratory considerations) apply to clusters which are still in a
non-steady state.  We must keep in mind, however, that dynamo effects
in clusters may be more complex than has been considered so far.

\section{Impact of the Field on the ICM}

To summarize:  the  observations suggest that microgauss fields
 extend throughout the cluster plasma.  Larger fields exist around
embedded radio sources;  it is not yet clear if they are
representative of the general diffuse ICM.  What are the consequences
of a such magnetic fields for the larger questions about the clusters? 

\subsection{Microphysics}

There has been quite a lot of discussion in the literature regarding
the effect of magnetic fields on transport processes in the ICM.  This
has come up most particularly in regard to thermal conduction (or its
suppression) in cooling-flow models;  and it is also relevant to
questions of relativistic particle diffusion in radio haloes.  The
detailed workings of particle transport depend on the detailed
structure of the magnetic field, particularly on small scales.  The 
literature tends to split into two camps:  those working with a
spectral representation of the turbulent magnetic field (\eg, Jokipii
1966),  and those working with a chaotic representation (\eg, Chandran
\& Cowley 1998).   The essence of both approaches involves determining
a characteristic, coherence, length for the turbulent field (if,
indeed, one exists).  It may be here that observations (such as the
structure of Faraday rotation images) can make connection with the
theory. 

\subsection{Structure}

A significant magnetic field in the center of the cluster will of
course affect the hydrostatic balance, and thus the structure, of the
ICM.  In a (quasi) static 
situation the field can provide an important part of the pressure
support.  This may be part of the answer to the discrepancy found
between cluster masses determined from gravitational lensing, and from
X-ray imaging (Miralda-Escud\'e \& Babul, 1995;  Loeb \& Mao 1994).
Just how general this result is does not seem clear yet, as only a few
clusters have been so analyzed both in X-rays and with lensing. It is also
worth noting that X-ray mass determinations will also be misleading if
the ICM is itself dynamic rather than in hydrostatic equilibrium (\eg,
Squires \etal, 1996).   The theoretical approach to this question will
probably lean heavily on numerical simulations of cluster evolution
(\eg, Garasi \& Eilek 1999, Roettiger 1999). 

\subsection{Dynamics}

A significant central magnetic field will also impact the dynamics of
the ICM.  In general, a magnetic field providing pressure support to
gas at the bottom of a potential well is unstable; it is energetically
favorable for the dense gas above to fall, and the magnetized, lower
density gas to rise.  On small scales, one would expect turbulence or
convection to be driven by magnetic bouyancy (\cf\ Balbus and Soker,
1989, also Zoabi, Soker \& Regev 1996, for applications to clusters).
On larger scales, one would expect a magnetically supported cluster
core to be subject to Parker-type instabilities (Parker, 1966).  In
either regime, the end effect should be the same -- we should expect
gas in the central region of a magnetized cluster to be disordered,
turbulent, and anything but static.

\begin{acknowledgements}
I am grateful to many of my colleagues for 
conversations and collaborations relating to magnetic fields in
clusters.  It is a particular pleasure to thank  Frazer Owen, Chris
Garasi, Ellen Zweibel, Greg Taylor, and 
Tracy Clarke for recent discussions. However, any errors or
misconceptions herein are my fault, not theirs.  This work was
partially supported by NSF grant AST-9720263.
\end{acknowledgements}

\end{document}